\documentclass[showpacs,prd,twocolumn,floatfix,
nofootinbib,superscriptaddress]{revtex4}
\usepackage{graphicx}

\newcommand{\be}{\begin{equation}}
\newcommand{\ee}{\end{equation}}
\newcommand{\bea}{\begin{eqnarray}}
\newcommand{\eea}{\end{eqnarray}}
\newcommand{\nl}{\nonumber \\}

\newcommand{\delv}{{\bf \nabla}}
\newcommand{\delvt}{\tilde{{\bf \nabla}}}

\newcommand{\Mbz}{{M_0}}
\newcommand{\delfour}{{\Delta^{(4)}}}
\newcommand{\delsq}{\Delta^{(2)}}

\newcommand{\Ev}{\tilde{{\bf E}}}
\newcommand{\Bv}{\tilde{{\bf B}}}

\newcommand{\sigmav}{\mbox{\boldmath$\sigma$}}

\begin{document}


\title{  B Meson Semileptonic Form Factors 
from Unquenched  Lattice QCD}

\author{Emel Gulez}
\author{Alan Gray}
\affiliation{Department of Physics,
The Ohio State University, Columbus, OH 43210, USA }

\author{Matthew Wingate }
\affiliation{Institute for Nuclear Theory, University of Washington,
Seattle, WA 98195-1550, USA }

\author{Christine T.\ H.\ Davies }
\affiliation{Department of Physics \& Astronomy,
University of Glasgow, Glasgow, G12 8QQ, UK }

\author{G. Peter Lepage }
\affiliation{Laboratory of Elementary Particle Physics,
Cornell University, Ithaca, NY 14853, USA}

\author{Junko Shigemitsu}
\affiliation{Department of Physics,
The Ohio State University, Columbus, OH 43210, USA }

\collaboration{HPQCD Collaboration}
\noaffiliation



\begin{abstract}
The semileptonic process, $B \longrightarrow \pi \, l \nu$, 
is studied via full QCD lattice simulations.
We use  unquenched 
gauge configurations generated by the MILC collaboration. These include 
the effect of vacuum polarization from three quark flavors: the $s$ quark 
and two very light flavors ($u/d$) of variable mass allowing extrapolations 
to the physical chiral limit.
 We employ Nonrelativistic QCD 
to simulate the $b$ quark and a highly improved staggered quark action 
for the light sea and valence quarks.
We calculate the form factors
$f_+(q^2)$  and $f_0(q^2)$ in the chiral limit for the range $16 \,
 {\rm GeV}^2 \leq q^2 < q^2_{max}$ and obtain  $
 \int_{16 GeV^2}^{q^2_{max}} [\,d\Gamma/d q^2\,] \,
dq^2 \; / \;|V_{ub}|^2 = 1.46(35) \, ps^{-1}$.
Combining this with a preliminary average by the Heavy Flavor Averaging Group 
(HFAG'05) of recent branching fraction data 
for exclusive B semileptonic decays from the BaBar, Belle and CLEO 
collaborations, leads to $|V_{ub}| = 4.22(30)(51) \times 10^{-3}$. \\
PLEASE NOTE APPENDIX B with an ERRATUM, to appear in Physical Review D, 
to the published version of this e-print (Phys.Rev.D 73, 074502 (2006)). 
Results for the form factor $f_+(q^2)$ in the chiral limit have changed
 significantly.  The last two sentences in 
this abstract should now read; \\
``We calculate the form factors
$f_+(q^2)$  and $f_0(q^2)$ in the chiral limit for the range $16 \,
 {\rm GeV}^2 \leq q^2 < q^2_{max}$ and obtain  $
 \int_{16 GeV^2}^{q^2_{max}} [\,d\Gamma/d q^2\,] \,
dq^2 \; / \;|V_{ub}|^2 = 2.07(57) \, ps^{-1}$.
Combining this with a preliminary average by the Heavy Flavor Averaging Group
(HFAG'05) of recent branching fraction data
for exclusive B semileptonic decays from the BaBar, Belle and CLEO
collaborations, leads to $|V_{ub}| = 3.55(25)(50) \times 10^{-3}$.''

\end{abstract}


\maketitle


\section{ Introduction}
A major achievement of the B factories in recent years has been 
the observation of CP violation in the neutral $B$ system \cite{babar1,
belle1}.
The emphasis since then has been on 
overconstraining the unitarity triangle and checking for consistency with,
 or deviations from, the three family Standard Model.
The goal is to independently measure not only the three angles 
but also the lengths of the sides of the triangle and thereby 
determine ($\overline{\rho},\overline{\eta}$), the apex of 
the unitarity triangle, in as many ways as possible.
 In order to fix the sides of the unitarity triangle,
the magnitudes of several Cabibbo-Kobayashi-Maskawa (CKM)
 matrix elements are required 
and the accuracy is currently 
limited mainly by theoretical uncertainties in two of them,
$|V_{td}|$ and $|V_{ub}|$. 
Theory input for these quantities involves hadronic matrix elements 
of several electroweak operators and these in turn require 
good control over nonperturbative QCD.  This article reports on 
significant recent progress in lattice QCD determinations of form 
factors relevant for the CKM matrix element $|V_{ub}|$.

$|V_{ub}|$ can be determined from studies of either inclusive or 
exclusive $B$ meson semileptonic decays.  The first determinations 
relied on inclusive measurements. However, recent impressive progress 
in measurements of branching fractions for exclusive decays by 
CLEO \cite{cleo1}, Belle \cite{belle2,belle3}
 and BaBar \cite{babar2,babar3,babar4}
 have started to make exclusive determinations 
 competitive \cite{nierste,stewart}. 
 In either approach errors are now dominated by 
theory errors and which method will eventually win out  depends
on how well the theoretical uncertainties in shape functions (for 
the inclusive approach) or form factors (in the case of the 
exclusive approach) can be brought under control. Lattice QCD 
provides a first principles nonperturbative QCD method for calculating 
form factors in exclusive semileptonic decays.  The first lattice 
calculations were carried out in the quenched approximation that 
ignored vacuum polarization effects \cite{ukqcd,ape,jlqcd,fermi,
d234}.  For some period these pioneering results were 
the only ones available and experimentalists, 
 e.g. the CLEO collaboration, used them to extract some of the earliest 
exclusive  $|V_{ub}|$ results \cite{cleo1}.
In the summer of 2004 the first 
preliminary unquenched results for the $B \rightarrow 
\pi \, l \nu$ form factors were presented by the Fermilab/MILC \cite{fermmilc}
 and 
HPQCD \cite{hpqcd} collaborations.  
These calculations employed the MILC collaboration 
$N_f = 2+1$ unquenched configurations, the most realistic gauge configurations 
to date with vacuum polarization from 2 flavors of very light quarks and 
from strange quarks \cite{milc1}.
 Furthermore, the good chiral properties of the 
improved staggered quark action used for the light sea and valence 
quarks, allowed for investigations much closer to the 
chiral limit than in the earlier calculations. 
Hence,  references \cite{fermmilc,hpqcd} constitute 
a major step forward in lattice QCD determinations of semileptonic 
form factors. These preliminary results have been 
incorporated by Belle and BaBar into their 
recent $B$ semileptonic analysis \cite{belle2,belle3,babar2,babar3,babar4} and 
used by other theorists in their extractions of $|V_{ub}|$ \cite{agrs,rh,bh}.

References \cite{fermmilc,hpqcd} are part of a growing 
list of recent lattice calculations 
that use the MILC unquenched configurations.  The creation of these 
 configurations became feasible on present day computers 
due to the development of highly improved staggered light quark 
actions \cite{asqtad}. 
There is one well known drawback of staggered actions, namely 
that each flavor comes in four types, called ``tastes''.
To simulate just 
one taste of sea quark per flavor a fourth root of the 
quark determinant is used and the validity of this procedure has not yet been 
rigorously proven.
All tests undertaken to date 
to study the fourth root procedure, however,  have led to 
encouraging results \cite{stgtests1}.
  There are no indications of problems or 
any deviations from continuum QCD behavior beyond small and expected
 discretisation errors that can be systematically improved upon. 
A recent review of the fourth root issue is given in \cite{stgtests2}. 
The MILC unquenched configurations and the 
heavy and light quark actions currently in use 
have also  been tested repeatedly by calculating a large
set of well measured quantities such as light hadron spectroscopy, light 
meson decay constants, quarkonium and B meson spectroscopy. Agreement 
between experiment and lattice calculations is found to be 
excellent within the few percent errors in the lattice results 
\cite{prl,milc2,lmass,upsilon,lhpc}.
 More impressively, 
these same gauge configurations have led to some recent predictions from 
Lattice QCD that have been confirmed subsequently by experiment
 \cite{mbc,fd,dform}.
 Most relevant 
for $B$ semileptonic decays, is the successful calculation by the 
Fermilab/MILC collaboration of $D$ meson semileptonic form factors
and the agreement of their $q^2$ dependence with experiment \cite{dform}.
All this should boost confidence in results for $B$ meson semileptonic decays 
that are now emerging from Lattice QCD calculations.

In this article we make several further improvements on the 
preliminary calculations of ref.\cite{hpqcd}
 and finalize our form factor results.
Some of these improvements have been reported on in \cite{emellat05}. 
Among the improvements, we have now included all the dimension 4 current 
corrections through ${\cal O}(\alpha_s/M)$ and ${\cal O}(a \, \alpha_s)$ 
 to the temporal and spatial components of the heavy-light 
electroweak currents relevant for $B \rightarrow \pi \, l \nu$ 
semileptonic decays. Our previous results included currents only at lowest 
order in $1/M$ and through ${\cal O}(\alpha_s)$.  We now have simulation 
data from several MILC ensembles, whereas in ref.\cite{hpqcd} results were at 
a single fixed light sea quark mass and only the valence quark 
mass was varied.  Another development since ref.\cite{hpqcd}
 is that we now employ 
the formulas of staggered chiral perturbation theory 
\cite{schpt1,schpt2,schpt3} simultaneously 
to both partially quenched and full QCD results to extrapolate in the 
light quark mass 
to the physical pion.  Previously we used linear chiral extrapolations to 
our partially quenched data.

So, the list of improvements in our semileptonic decay studies 
 since the work of ref.\cite{hpqcd} is quite substantial. 
Nevertheless, we will see that changes in 
the final results for the form factors 
$f_+(q^2)$ and $f_0(q^2)$ are almost negligible. The individual 
contribution from each higher dimension current correction 
is found to be small, 
mainly due to small one-loop matching coefficients. There is also some 
cancellation between the many terms.  The difference between linear 
chiral extrapolations of our previous partially quenched data
 and results from staggered chiral perturbation 
theory fits to the new full data set also turns out to be a small effect. 
  With all the improvements now in place, our semileptonic 
form factor calculations are at the same level as recent $B$ meson 
decay constant determinations by the HPQCD collaboration \cite{fb,fbs}. 
  Those latter 
calculations are crucial for determinations of the CKM matrix 
elements $|V_{td}|$ and $|V_{ts}|$.

In the next section we provide some details of the simulation 
parameters and of the lattice actions employed.  We also summarize 
formulas for the relevant form factors. 
In section III. we discuss matching of the lattice heavy-light currents 
used in our simulations to their continuum QCD counter parts. 
One-loop matching coefficients for 
the temporal components $V_0$ and $A_0$ were published in ref.
\cite{pert1} and have been 
used already in our decay constant determinations \cite{fb,fbs}.
Matching of the spatial components $V_k$ was completed as part of 
the current project.  Section IV. describes how we extract 
the form factors $f_\parallel$ and $f_\perp$ from numerical simulations.
The more commonly used form factors $f_+$ and $f_0$ can be expressed 
as simple linear combinations of $f_\parallel$ and $f_\perp$.
Section V. focuses on chiral extrapolations of $f_\parallel$ and $f_\perp$.
In section VI. we summarize our final results for $f_+(q^2)$ and 
$f_0(q^2)$ in the physical chiral limit. We also present Tables 
of partially integrated differential decay rates (divided by $|V_{ub}|^2$) 
in several $q^2$ bins.  In section VII. we combine the results of 
section VI. with experimental data on $B \rightarrow \pi \, l \nu$ 
branching fractions to estimate $|V_{ub}|$.  We then conclude with 
a summary section.

\section{Simulation Details and Form Factor Formulas}

\begin{table}
\begin{center}
\begin{tabular}{|c|c|c|c|c|c|}
\hline
$N_s^3 \times N_t$  &  $a^{-1}(GeV)$ & $N_{conf}$ & $u_0 \, a m_f$ &
$u_0 \, a m_q$ & $m_q/m_s$\\
\hline
\hline
 Coarse  &&&&& \\
$24^3 \times 64$  &  1.623(32)$^\dagger$  & 399 & 0.005 &  0.005 & 0.125   \\
$20^3 \times 64$    &  1.622(32)$^\dagger$  &  397 & 0.007 & 0.007 &0.175  \\
$20^3 \times 64$    &  1.596(30)  &  568 & 0.010 & 0.005 & 0.125 \\
          &&&                               & 0.010 & 0.250  \\
          &&&                               & 0.020 &0.500   \\
 $20^3 \times 64$   &   1.605(29)   & 486  & 0.020 & 0.020 & 0.500  \\
\hline
Fine  &&&&&  \\
$28^3 \times 96$   &   2.258(32)   &  465  & 0.0062 & 0.0062 & 0.200 \\
$28^3 \times 96$  &   2.312(31)   &  496  & 0.0124 & 0.0124 &0.400 \\
\hline
\end{tabular}
\caption{Simulation Details. $m_f$ ($m_q$) denotes sea (valence) quark masses 
and $m_s$ is the strange quark mass. 
 A $\dagger$  means the lattice spacing was determined through $r_1$.
 For all other ensembles
the $\Upsilon$ 2S-1S splitting was used.
$u_0 = [plaq]^{1/4}$ is the link variable that enters into 
the MILC collaboration's convention for normalisation of quark masses.
}
\end{center}
\end{table}

Most of our simulations were performed on the MILC collaboration ``coarse'' 
ensembles with lattice spacing around $0.12$fm and with light sea quark masses 
in units of the strange quark mass, $m_f/m_s$, ranging between 0.125 and 0.5.
We have also carried out some checks using MILC lattices with 
finer lattice spacings.
 We refer to the original papers by the MILC collaboration 
for details of how their lattices were generated \cite{milc1}.
 Some  specifications for the ensembles are given in Table I. 
The highly improved staggered action, the AsqTad action, is used for 
both the sea and valence light quarks. The leading discretisation errors in 
this formalism are 
${\cal O}(a^2 \, \alpha_s)$.  To simulate the $b$ quark inside the $B$ meson 
we use the same highly improved
 nonrelativistic (NRQCD) action  employed in recent 
studies of the $\Upsilon$ system on the same MILC ensembles 
\cite{upsilon,ups2}. 
 In terms of
the two-component Pauli spinor $\Phi$ one has,
\bea
 \label{nrqcdact}
&& {\cal L}_{\rm NRQCD}  =
\sum_x \Bigg\{  {\Phi}^\dagger_t \Phi_t -
 {\Phi}^\dagger_t
\left(1 \!-\!\frac{a \delta H}{2}\right)_t
 \left(1\!-\!\frac{aH_0}{2n}\right)^{n}_t \nonumber \\
& \times & \frac{1}{u_0} \,
 U^\dagger_t(t-1)
 \left(1\!-\!\frac{aH_0}{2n}\right)^{n}_{t-1}
\left(1\!-\!\frac{a\delta H}{2}\right)_{t-1} \Phi_{t-1} \Bigg\} \, ,
 \eea
where 
$n$ is a stability parameter 
introduced to control high momentum modes of the $b$ propagator 
\cite{upsilon}.  
 $H_0$ is the nonrelativistic kinetic energy operator,
 \be
 H_0 = - {\delsq\over2\Mbz} \, ,
 \ee
and $\delta H$ includes relativistic and finite-lattice-spacing
corrections,
 \begin{eqnarray}
\delta H
&=& - c_1\,\frac{(\delsq)^2}{8\Mbz^3}
+ c_2\,\frac{i g}{8\Mbz^2}\left(\delv\cdot\Ev - \Ev\cdot\delv\right) \nl
& &
 - c_3\,\frac{g}{8\Mbz^2} \sigmav\cdot(\delvt\times\Ev - \Ev\times\delvt)\nl
& & - c_4\,\frac{g}{2\Mbz}\,\sigmav\cdot\Bv
  + c_5\,\frac{a^2\delfour}{24\Mbz}  - c_6\,\frac{a(\delsq)^2}
{16n\Mbz^2} \, .
\label{deltaH}
\end{eqnarray}
 $\Mbz$ is the bare $b$-quark mass,
$\delsq$ the lattice Laplacian, $\delv$ the symmetric lattice derivative 
and $\delfour$ the lattice discretization of the continuum $\sum_i D^4_i$. 
Expressions for the improved $\Ev$ and $\Bv$ fields are given in 
\cite{upsilon}.
All derivatives are tadpole improved. 
As in \cite{upsilon} we set the $c_i$'s to their tree level value, 
$c_i = 1$.  These coefficients will be modified by radiative 
corrections at higher order. For heavy-light systems the most important 
radiative corrections will be to $c_4$. All other $c_i$'s are multiplied 
by additional factors of $1/M$ or $a$. 
Based on the discussions in 
\cite{upsilon} we estimate the effect from radiative corrections
 to $c_4$ on
$B$ semileptonic form factor calculations to be less than 1\%.

Just as in continuum QCD, our lattice actions include a small number 
of parameters that can only be fixed via experimental input. These are 
the bare quark mass parameters and the scale (or coupling). 
For the present case of heavy-light simulations the action parameters 
have already been fixed for us via simulations of quarkonium and 
light quark systems using the same MILC configurations. 
We use lattice spacings determined 
 by the $\Upsilon$ 2S-1S splitting \cite{upsilon}. 
 For two of the MILC ensembles (denoted by a $\dagger$ in Table I),
  $\Upsilon$ simulations have not yet been carried out. On the other
  hand, the MILC collaboration has measured the heavy quark potential 
 parameter $r_1/a$ (in lattice units)
 for these ensembles \cite{milc2}.  In reference [28], using
 MILC ensembles on which both $r_1/a$ and the $\Upsilon$ 2S-1S splittings
 were calculated, we could use the experimental 2S-1S splitting
 to determine $r_1 = 0.321(5)$fm.  This physical value for $r_1$
 was combined with the MILC collaboration $r_1/a$ to fix the scale
 for the above two ensembles.  
 The $b$ quark mass is also fixed by our $\Upsilon$ studies.  Studies 
of pions and kaons have fixed the $u$ and $d$ masses (which we take 
to be equal to each other) and the $s$ quark mass respectively
 \cite{milc2,lmass}.  Hence by the 
time one gets to the $B$ system  there are no adjustable parameters left 
in our QCD action.

To study the process $B \rightarrow \pi \, l \nu$, one needs to 
evaluate the matrix element of the charged electroweak  current 
between the $B$ and the $\pi$ states, $\langle \pi | (V - A)^\mu | B \rangle$. 
Only the vector current $V^\mu$ contributes to the pseudoscalar-to-pseudoscalar
 amplitude and the matrix 
element can be written in terms of two form factors $f_+(q^2)$ and 
$f_0(q^2)$. These depend only on the square of the momentum transfered 
between the $B$ and the $\pi$, $q^\mu = p_B^\mu - p_\pi^\mu$.
\begin{eqnarray}
\label{f0plus}
\langle \pi(p_\pi)| V^\mu| B(p_B) \rangle &=&
f_+(q^2) 
\, \left[ p_B^\mu + p_\pi^\mu - \frac{M_B^2-m_\pi^2}{q^2}
\, q^\mu \right] \nonumber \\
 &+& f_0(q^2)
 \; \frac{M_B^2 - m_\pi^2}{q^2} \; q^\mu .
\end{eqnarray}
If one neglects the mass of the charged lepton in the final state, only 
the form factor $f_+(q^2)$  contributes to the 
decay rate $\Gamma(B \rightarrow \pi \, l \nu)$. 
 Nevertheless, it is useful to keep track of the form factor 
$f_0(q^2)$ as well since, as we shall see, it helps in our interpolations and 
extrapolations of simulation data.  In our data analysis another pair 
of form factors, $f_\parallel$ and $f_\perp$, turn out to be more convenient.
\be
\label{fpaperp}
\langle \pi(p_\pi)| V^\mu| B(p_B) \rangle =
 \sqrt{2 M_B} \,[v^\mu f_\parallel\,
 + \, p^\mu_\perp  f_\perp ] ,
\ee
with
\be
v^\mu = \frac{p_B^\mu}{M_B} \qquad , \qquad p_\perp^\mu
=p_\pi^\mu - (p_\pi \cdot v) \, v^\mu .
\ee
In the $B$ rest frame (in this article we only consider $B$ mesons 
decaying at rest) the temporal and spatial parts of (\ref{fpaperp}) 
become,
\begin{eqnarray}
\label{fpaperp0}
\langle \pi| V^0 |B \rangle &=&
 \sqrt{2 M_B} \,       f_\parallel \nonumber \\
\langle \pi| V^k |B \rangle &=&
 \sqrt{2 M_B} \,p^k_\pi \, f_\perp .
\end{eqnarray}
Hence, one sees that one can separately determine $f_\parallel$ or 
$f_\perp$  simply by looking at either the temporal or spatial 
component of $V^\mu$. These two form factors have the additional 
advantage that they have simpler HQET scaling properties and 
chiral perturbation theory is carried out in terms of them rather 
than for $f_+$ and $f_0$.  After carrying out the chiral extrapolations 
for $f_\parallel$ and $f_\perp$, we convert back to obtain $f_+$ and 
$f_0$ for the physical theory using,
\be
f_+ = \frac{1}{\sqrt{2 M_B}} \, f_\parallel + \frac{1}{\sqrt{2 M_B}} \,
(M_B - E_\pi) \, f_\perp
\ee
\be
f_0 = \frac{\sqrt{2 M_B}}{(M_B^2 - m_\pi^2)}\, [ (M_B- E_\pi) f_\parallel
+ (E_\pi^2 - m_\pi^2) f_\perp ] ,
\ee
where $E_\pi$ is the pion energy in the $B$ rest frame. 
From these formulas one sees that $f_+$ will be dominated by
$f_\perp$, i.e. by the matrix element of $V_k$, and $f_0$ by $f_\parallel$ or
the matrix element of $V_0$.

Our goal is to evaluate the hadronic matrix elements 
$\langle \pi | V^0|B\rangle$ and 
$\langle \pi | V^k|B\rangle$ via lattice simulations.  There are several 
steps in the calculation.  First, one must relate the continuum electroweak 
currents,
 $V^0$ and $V^k$, to lattice operators written in terms of the heavy and light 
quark fields in our lattice actions.  In the second step the matrix 
elements of these lattice current operators must be evaluated numerically 
and the relevant amplitude, i.e. the matrix element between the ground state 
$B$ meson and the ground state pion with appropriate momenta must 
be extracted.  This will give us, via eqns.(\ref{fpaperp0}), the 
form factors $f_\parallel$ and $f_\perp$ as functions of the light quark 
mass and the pion momentum.  Finally in step 3 these numerical results
 must be extrapolated to the physical pion. 
In the next three sections 
we describe each of these three steps in turn.

\section{Matching of Heavy-Light Currents }
Matching of heavy-light currents between continuum QCD and a lattice 
effective theory with two-component nonrelativistic heavy quark 
fields $\Phi$ and four-component light quarks $q(x)$ is discussed in 
ref.\cite{pert2}. Since  staggered light quarks 
can be written in terms of four-component ``naive'' AsqTad quark fields 
the formalism developed there carries over unchanged to the present 
calculation.  Introducing also a four-component notation for the heavy
field, $Q(x) \equiv (\Phi, 0)$, one finds that through 
 ${\cal O}(\alpha_s \, \Lambda_{QCD}/M 
\; , \; \alpha_s/(aM) \; , \; \alpha_s \, a \, \Lambda_{QCD})$ the 
following current operators in the effective theory are required.

\underline{ temporal} :
\begin{eqnarray}
\label{j0}
 J^{(0)}_{0}(x) & = & \bar q(x) \,\Gamma_0\, Q(x), \nonumber \\
 J^{(1)}_{0}(x) & = & \frac{-1}{2M_0} \bar q(x)
    \,\Gamma_0\,\mbox{\boldmath$\gamma\!\cdot\!\nabla$} \, Q(x),\nonumber \\
 J^{(2)}_{0}(x) & = & \frac{-1}{2M_0}  \bar q(x)
    \,\mbox{\boldmath$\gamma\!\cdot\!\overleftarrow{\nabla}$}
    \,\gamma_0\ \Gamma_0\, Q(x). 
\end{eqnarray}

\underline{ spatial } :
\begin{eqnarray}
\label{jk}
 J^{(0)}_{k}(x) & = & \bar  q(x) \,\Gamma_k\, Q(x),\nonumber \\
 J^{(1)}_{k}(x) & = & \frac{-1}{2M_0} \bar q(x)
    \,\Gamma_k\,\mbox{\boldmath$\gamma\!\cdot\!\nabla$} \, Q(x),\nonumber \\
 J^{(2)}_{k}(x) & = & \frac{-1}{2M_0} \bar q(x)
    \,\mbox{\boldmath$\gamma\!\cdot\!\overleftarrow{\nabla}$}
    \,\gamma_0\ \Gamma_k\, Q(x), \nonumber  \\
 J^{(3)}_{k}(x) & = & \frac{-1}{2M_0} \bar q(x)\,  \nabla_k
\, Q(x) ,  \nonumber \\
 J^{(4)}_{k}(x) & = & \frac{1}{2M_0} \bar q(x)
    \,\overleftarrow{\nabla}_k  \, Q(x), 
\end{eqnarray}
where $\Gamma_\mu$ can be either $\gamma_\mu$ or $\gamma_5 \gamma_\mu$, 
and $M_0$ is the bare heavy quark mass in the NRQCD action.
One sees that there are two dimension 4 current corrections to the 
temporal component and four such corrections to the spatial components. 
To the order that we are working, one has 
\begin{eqnarray}
\label{v0}
\langle V_0 \rangle  &=& ( 1 + \alpha_s \, 
\tilde{\rho}^{(0)}_0)\,\langle J^{(0)}_0 \rangle + \nonumber \\
 & & (1 + \alpha_s   \,  \rho^{(1)}_0) \, \langle
J^{(1),sub}_0 \rangle + \alpha_s  \,
 \rho^{(2)}_0 \, \langle J^{(2)}_0 \rangle
\end{eqnarray}
and
\begin{eqnarray}
\label{vk}
\langle V_k \rangle  &=& ( 1 + \alpha_s \, 
\tilde{\rho}^{(0)}_k)\,\langle J^{(0)}_k \rangle + \nonumber \\
 & & (1 + \alpha_s   \,  \rho^{(1)}_k) \, \langle
J^{(1),sub}_k \rangle 
+ \alpha_s  \,
 \rho^{(2)}_k \, \langle J^{(2)}_k \rangle
\nonumber  \\
 & & + \alpha_s  \,
 \rho^{(3)}_k \, \langle J^{(3)}_k \rangle
+ \alpha_s  \,
 \rho^{(4)}_k \, \langle J^{(4)}_k \rangle.
 \end{eqnarray}
We introduce the combination 
$ J_\mu^{(1),sub} = J_\mu^{(1)} - \alpha_s \,   \zeta_{10,\mu}
J_\mu^{(0)}$. This subtracts out power law contributions to 
the matrix elements of the higher dimension operator $J^{(1)}_\mu$ 
 through ${\cal  O}(\alpha_s/(aM))$ \cite{powerlaw}. 
  $J_\mu^{(1)}$ enters the 
matching already at tree level and after the subtraction 
 one is left with the physical 
${\cal O}(\Lambda_{QCD}/M)$  contribution that is a relativistic correction 
to the leading order term.  
Power law subtractions of the other dimension 4 current corrections come in as 
${\cal O}(\alpha_s^2/(aM))$ effects and are only partially included here. 
The one-loop coefficients $\rho_\mu^{(j)}$ and $\zeta_{10,\mu}$ for 
$\mu = 0$ are given in ref.\cite{pert1}.
  The results for $\mu = k$ have not been 
published before and are summarized in Table II.  In ref.\cite{hpqcd} only the 
contributions from the first lines in eqns.(\ref{v0}) and (\ref{vk}) were 
taken into account, i.e. $J_0^{(0)}$ and $J_k^{(0)}$ matched through 
${\cal O}(\alpha_s)$.

\begin{table}
\begin{center}
\begin{tabular}{|c|c|c|c|c|c|c|c|}
\hline
$aM_0$ & n & $\tilde{\rho}_k^{(0)}$ & $\rho_k^{(1)}$ & $\rho_k^{(2)}$ & 
$\rho_k^{(3)}$ & $\rho_k^{(4)}$ & $\zeta_{10,k}$ \\
\hline
\hline
4.00  &  2  & 0.256 & 0.484(3) & 0.340(6)& 0.244(3)& -0.137(3)& 0.041 \\
2.80  &  2  & 0.270 & 0.349(3) & 0.169(6) & 0.218(4) & -0.029(4) & 0.055 \\
1.95  &  2  & 0.332 & 0.232(3) & 0.121(8) & 0.161(4)& 0.063(3)&  0.073 \\
\hline
\end{tabular}
\caption{ Matching coefficients for the spatial currents $V_k$. 
 Where errors are not indicated explicitly, 
they are of order one or less in the last digit.
$aM_0$ is the bare heavy quark mass in lattice units and $n$ a parameter 
in the NRQCD action. 
The three selected values for $aM_0$ correspond to the $b$ quark on the 
MILC extra-coarse, coarse and fine lattices respectively \cite{upsilon}.
}
\end{center}
\end{table}

\begin{figure}
\includegraphics[width=8.0cm,height=6.0cm]{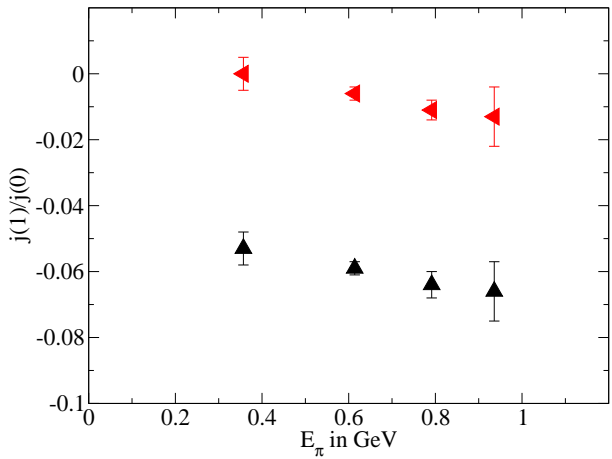}
\caption{  The ratio $\langle J_0^{(1)}\rangle / \langle J_0^{(0)} \rangle$ 
for one ensemble ($u_0 am_f = u_0 am_q = 0.01$)
 versus the pion energy $E_\pi$.  The lower points are before power law 
subtraction and the upper points after power law subtraction (i.e. 
$\langle J_0^{(1),sub}\rangle / \langle J_0^{(0)} \rangle$ ).
 }
\end{figure}

\begin{figure}
\includegraphics[width=8.0cm,height=6.0cm]{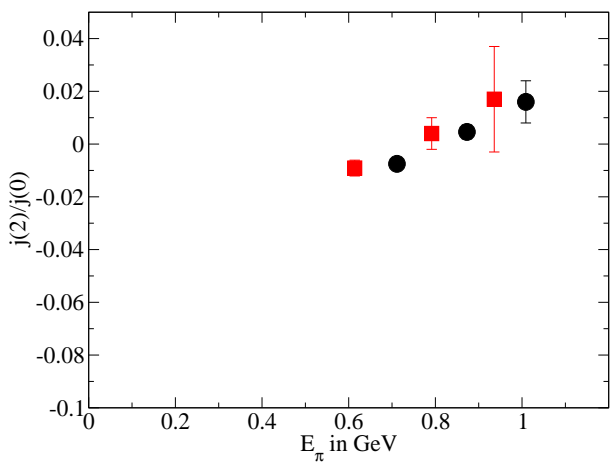}
\caption{  The ratio $\langle J_k^{(2)}\rangle / \langle J_k^{(0)} \rangle$ 
for two ensembles versus the pion energy $E_\pi$. Squares are for 
$u_0 am_f = 0.01$ and circles for $u_0 am_f = 0.02$.
 }
\end{figure}

\begin{figure}
\includegraphics[width=8.0cm,height=6.0cm]{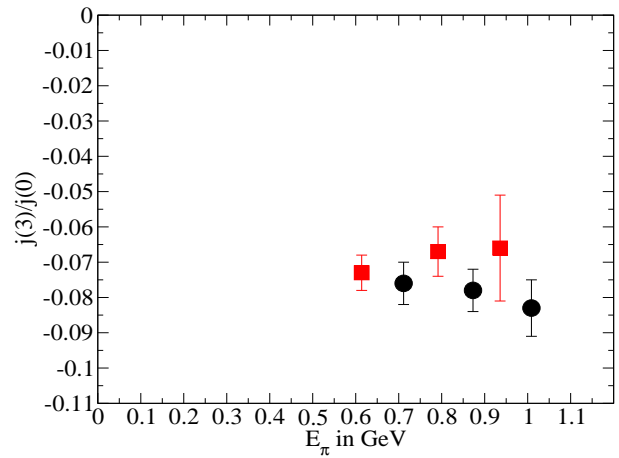}
\caption{ Same as Fig.2 
for  $\langle J_k^{(3)}\rangle / \langle J_k^{(0)} \rangle$ 
 }
\end{figure}

\begin{figure}
\includegraphics[width=8.0cm,height=6.0cm]{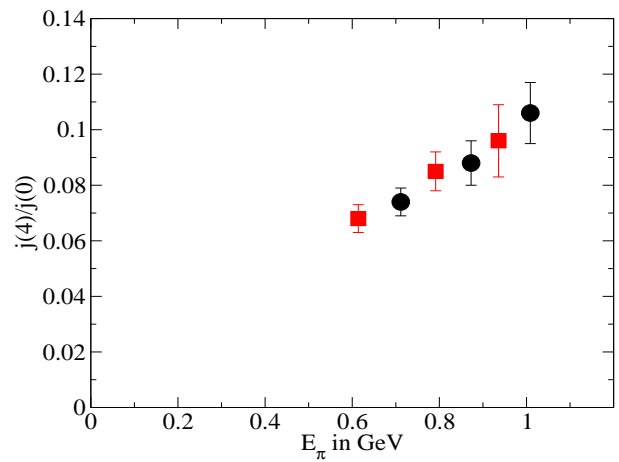}
\caption{ Same as Fig.2 
for  $\langle J_k^{(4)}\rangle / \langle J_k^{(0)} \rangle$ 
 }
\end{figure}

\vspace{.1in}
As mentioned in the introduction, the effects of all the dimension 4 
current corrections turn out to be very small. In Fig.1 we show results for 
$\langle J_0^{(1)}\rangle / \langle J_0^{(0)}\rangle $ for one of our
ensembles  with and without 
the power law subtraction. One sees that although the unsubtracted 
$\langle J_0^{(1)}\rangle / \langle J_0^{(0)}\rangle $ is at the 
$\sim6$\% level, the physical 
$\langle J_0^{(1),sub}\rangle / \langle J_0^{(0)}\rangle $ is
$\leq 1$\%.  In Figs.2 - 4 we give  further examples of 
$\langle J_k^{(j)}\rangle / \langle J_k^{(0)}\rangle $ for $j > 1$. 
These get multiplied 
by factors of $(\rho_k^{(j)} \alpha_s)$ in eq.(\ref{vk}). Using 
 $\alpha_s \approx \alpha_V(2/a) = 0.32$ \cite{alphas} and Table II, 
one finds $\rho \,\alpha_s$ factors 
between 0.01 and 0.11, which leads to contributions from 
higher order currents that are at most 1\%. For instance the largest current 
correction is $J_k^{(4)}$ (Fig.4), but $\rho_k^{(4)} \, \alpha_s
 = -0.029 \, \alpha_s = -0.0009$, and the contribution from this current 
is negligible. In comparing Figs.1-4 one sees that 
the size of matrix elements grows with the pion energy $E_\pi$ for 
$J_k^{(2)}$ and $J_k^{(4)}$ and seems much less sensitive to $E_\pi$ 
for the other two currents.  This reflects the fact that 
$J_k^{(2)}$ and $J_k^{(4)}$ have derivatives acting on the 
light quark field that is part of the final state pion and therefore 
knows about its momentum.

\section{Simulation Results for Form Factors $f_\parallel$ and $f_\perp$}
The starting point for calculations of the hadronic matrix elements 
$\langle \pi | J_\mu^{(j)} | B \rangle$ is the 3-point correlator, 
\begin{eqnarray}
\label{threepnt}
& & C^{(3)}(\vec{p}_\pi, \vec{p}_B, t, T_B)   =   \nonumber \\
 & & \sum_{\vec{z}} \sum_{\vec{y}}  \langle \Phi_\pi(0)
J_\mu^{(j)}(\vec{z},t) \Phi^\dagger_B(\vec{y},-T_B)  \rangle 
\, e^{i\vec{p}_B\cdot \vec{y}} \, e^{i(\vec{p}_\pi - \vec{p}_B)\cdot \vec{z}},
\nonumber \\
\end{eqnarray}
where 
$\Phi_B$ and $\Phi_\pi$ are interpolating operators for the $B$ meson and 
the pion respectively. All results here have 
the $B$ meson three momentum, $\vec{p}_B$, set equal to zero.  For simplicity, 
 the pion operator $\Phi_\pi$ was always placed at the origin. 
The $B$ meson was then created at time slice $- T_B$ and the electroweak 
current, $J_\mu^{(k)}$, that converts the $b$ quark into an $u$ quark was 
inserted at times $0 \geq t \geq -T_B$.  We have also simulated the 
time-reversed process, which then has the electroweak current 
inserted between $ +T_B \geq t \geq 0 $ and $\Phi_B$ acting on time slice 
$+T_B$. 
 By looking at both forward and 
time-reversed processes and verifying that they lead to consistent 
results (within statistical errors), we were able to increase statistics and 
at the same time provide some check on our codes.  For most of our simulations 
we used $T_B/a = 16$ on the coarse MILC lattices and 
$T_B/a = 24$ on the fine lattices.  On one of the coarse lattices we 
also ran with $T_B/a = 20$ and verified that results for form factors 
were independent of $T_B$.  Making $T_B$ too large is not helpful since 
statistical errors grow with $T_B$.  On the other hand making $T_B$ too 
small limits the number of data points available and gives us less flexibility 
in our fits.

In constructing the interpolating operators $\Phi_B$ and $\Phi_\pi$ 
we have found it convenient, just as in the currents of eqs.(\ref{j0}) and 
(\ref{jk}), to work with four component naive fields $q(x)$ 
rather than one component staggered fields $\chi(x)$. Hence in 
eq.(\ref{threepnt}) we use,
\be
\label{phis}
\Phi_B^\dagger = \overline{Q} \gamma_5 q \qquad , \qquad \Phi_\pi =
\overline{q} \gamma_5 q .
\ee
The relation between naive and staggered fermion propagators 
is given by \cite{stagghl},
\be
\label{nvestagg}
 G_q(x,y) = \Omega(x) \, \Omega^\dagger (y) \, G_\chi(x,y) 
\ee
with
\be
\Omega(x) = \prod^3_{\mu=0 } (\gamma_\mu)^{x_\mu} .
\ee
In our simulations we first calculate staggered propagators $G_\chi$, 
since they are cheaper, and then convert to naive propagators $G_q$ 
using eq.(\ref{nvestagg}) before evaluating  3-point correlators.
  The naive AsqTad theory has 16 tastes of quark 
per flavor and hence one could form 16 different heavy-light pseudoscalar 
boundstates. However, as discussed in ref.\cite{stagghl} these 
exactly degenerate 16 $B$ mesons 
do not mix and the 2-point correlator $\langle \Phi_B(x) \Phi_B^\dagger
(y) \rangle$ for instance, receives contributions from only one of these 
possible $B$ mesons.  Similarly one can argue that in eq.(\ref{threepnt}) 
only one type of $B$ meson is involved and that it connects to only 
one of the 16 true Goldstone pions of the naive light quark theory 
(out of the total number of 256 pions).  In other words both 
eq.(\ref{threepnt}) and the $\Phi_B$ two-point correlator have the same 
normalization as in a theory with undoubled light and heavy quarks, 
such as continuum QCD.  The one correlator where adjustment of 
normalization is required is the $\Phi_\pi - \Phi_\pi$ correlator. 
Using naive fields as in eq.(\ref{phis}) brings in an extra factor 
of 16 due to the trace over a $16 \times 16$ taste matrix and this factor of 
16 must be divided out. 
If one 
 works with conventional staggered light quarks one would have an 
extra factor of 4 rather than 16 relative to a pion correlator in a 
theory with undoubled fermions.

To extract the matrix elements 
$\langle \pi | J_\mu^{(j)}(t) | B \rangle$, the 3-point correlators must be 
fitted to:
\begin{eqnarray}
&& C^{(3)}(\vec{p}_\pi, \vec{p}_B, t, T_B)   \rightarrow 
  \sum_{k=0}^{N_\pi-1} \sum_{l=0}^{N_B-1} \nonumber \\
&&  (-1)^{k*t} \, (-1)^{l*(T_B-t)}
\,  A_{\mu,lk}^{(j)}
\,  e^{-E_\pi^{(k)} t} \, e^{ -E_B^{(l)} (T_B-t)}.  \nonumber \\
\end{eqnarray}
With this ansatz every second exponential 
($k$ or $l$ odd) corresponds to an 
oscillatory (in time) contribution to the correlator, a characteristic 
feature of staggered fermions.
We use Bayesian fitting methods \cite{bayse} and 
in most of our fits we kept 
$N_\pi = 1$ and let $N_B$ vary between $3 \sim 9$.  In order to avoid 
contamination from excited pions we dropped $5$ to $8$  points 
close to the pion source and made certain that within errors
 fit results did not depend on the number of points omitted.
 We have also tried some fits with 
$N_\pi = 2$ or $3$. For some light quark masses and pion momenta 
one obtained results consistent with the $N_\pi = 1$ fits, in other cases, 
however, it was not possible to get very stable (with respect to $N_B$) 
fits. Hence, for our final results we rely on the $N_\pi = 1$ fits. 
Since oscillatory contributions are much more significant in the 
$B$ channel than for pions, it was more important to allow $N_B$ to increase 
at fixed $N_\pi$ rather than the other way around. 
For priors in our Bayesian fits we use central values corresponding 
to energy splittings of about 400MeV allowing, however for large 
$\sim$100\% widths. As priors for the amplitudes we allow for ranges 
typically between about $-10 \times A_{00}$ and $+10 \times A_{00}$, 
where $A_{00}$ is the groundstate amplitude. Fit results for 
groundstate energies and amplitudes were very insensitive to 
 choices for priors.  
In Figs.5 \& 6 we show examples of fit results for 
$\langle J^{(0)}_\mu (t) \rangle$
 for pion momentum $\frac{2 \pi}{N_s a}
(0,1,1)$ on one of the coarse ensembles.  For comparison we show in 
Fig.7 results for a fit to the $B$ correlator that was done simultaneously 
with the fit to the 3-point correlator.

\begin{figure}
\includegraphics[width=8.0cm,height=8.5cm]{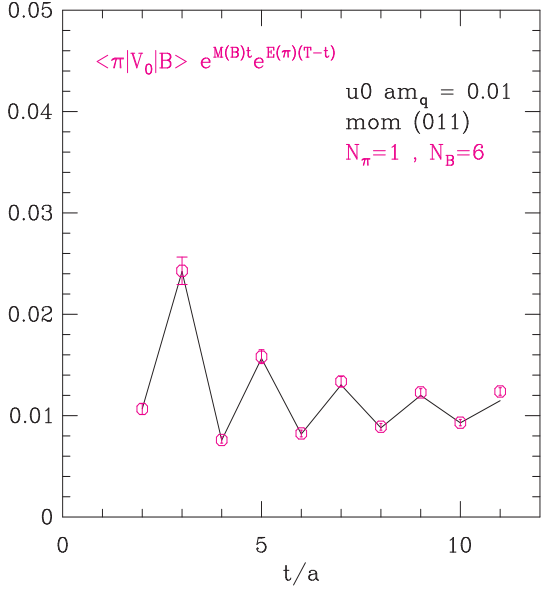}
\caption{ Fit  result for $\langle  J^{(0)}_0(t) \rangle 
\times e^{M_Bt} \times e^{E_\pi(T-t)}$ 
versus $t$ for 
pion momentum $(0,1,1)\frac{ 2 \pi}{N_s a}$.  The 
horizontal time axis has been rearranged so that the $B$ meson source 
is at $t/a = 1$ and the pion source at $t/a = T = 16$. 
 }
\end{figure}

\begin{figure}
\includegraphics[width=8.0cm,height=8.5cm]{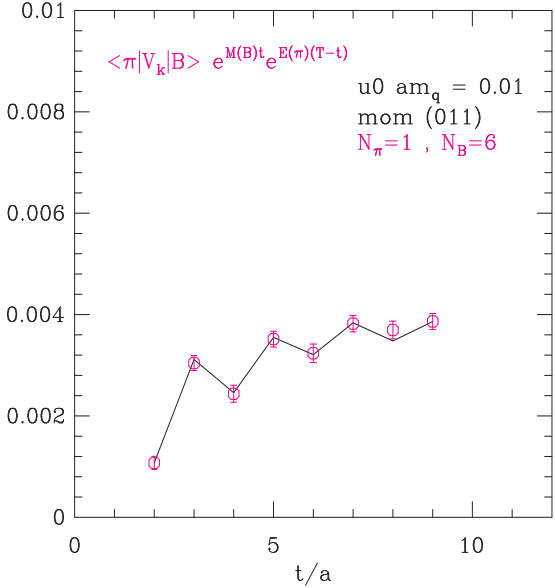}
\caption{ Same as Fig.5 but for $\langle  J^{(0)}_k(t)  \rangle$. 
 }
\end{figure}

\begin{figure}
\includegraphics[width=8.0cm,height=8.5cm]{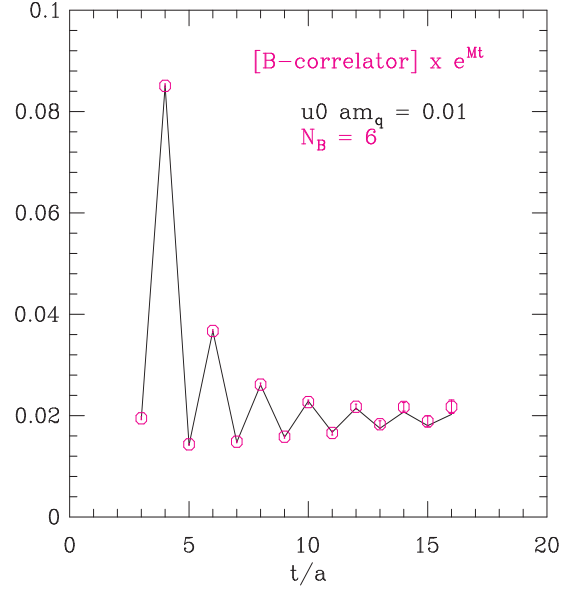}
\caption{ Fit results for the $B$ meson correlator
 }
\end{figure}

The main goal in all our fits is to extract the 
 ground state amplitudes $A_{\mu,00}^{(j)}$
which lead
directly to the form factors $f_\parallel$ and $f_\perp$ via
eq.(\ref{fpaperp0}).
\begin{eqnarray}
\label{fpaperp1}
f_{\parallel} & = & \frac{ A_{00}(V_0)}
{\sqrt{\zeta_\pi \zeta_B}}
\, \sqrt{2 E_\pi}   \nonumber \\
f_{\perp} & = & \frac{ A_{00}(V_k)}
{\sqrt{\zeta_\pi \zeta_B}
\;p_\pi^k }
\, \sqrt{2 E_\pi} 
\end{eqnarray}
Here $A_{00}(V_\mu)$ includes contributions from all currents with 
appropriate matching factors as dictated by eqns.(\ref{v0}) and (\ref{vk}). 
$\zeta_\pi$ and $\zeta_B$ are the ground state amplitudes from 
the pion and $B$ meson correlators, respectively.  $\zeta_\pi$ is correctly 
normalized as in continuum QCD. 
Results for the form factors $f_\parallel$ and $f_\perp$ for 
the different ensembles and pion momenta are summarized in Tables III 
\& IV.

\begin{table}
\begin{center}
\begin{tabular}{|c|c|c|c|c|c|}
\hline
$u_0 a m_f$ & $u_0 am_q$ & $p_\pi=(000)$ & $ (001)$  & 
$ (011)$ & $ (111)$\\
\hline
\hline
0.005  & 0.005  &1.486(34) &1.326(36) &1.221(38) & 1.127(78)  \\
0.007  & 0.007  &1.538(32) &1.328(33) &1.185(49) & 1.013(115)  \\
0.010  & 0.005  &1.582(37) &1.322(37) &1.201(60) & 1.053(121) \\
0.010  & 0.010  &1.584(44) &1.322(39) &1.212(52) & 1.064(138) \\
0.010  & 0.020  &1.581(30) &1.372(29) &1.253(37) & 1.117(72) \\
0.020  & 0.020  &1.508(48) &1.378(43) &1.264(47) & 1.121(83) \\
\hline
\end{tabular}
\caption{ Results for the form factor $ f_\parallel$ in 
GeV$^{1/2}$ from coarse MILC lattices. Errors are statistical errors 
coming from Bayesian bootstrap fits.
}
\end{center}
\end{table}

\begin{table}
\begin{center}
\begin{tabular}{|c|c|c|c|c|}
\hline
$u_0 a m_f$ & $u_0 am_q$  & $p_\pi = (001)$  & 
$ (011)$ & $ (111)$\\
\hline
\hline
0.005  & 0.005  &1.543(205) &0.703(44) & 0.515(39)  \\
0.007  & 0.007  &1.082(31) &0.606(31) & 0.433(48)  \\
0.010  & 0.005  &1.128(37) &0.662(36) & 0.413(38)  \\
0.010  & 0.010  &1.235(90) &0.657(37) & 0.449(43)  \\
0.010  & 0.020  &1.029(21) &0.611(19) & 0.423(23)  \\
0.020  & 0.020  &1.097(140) &0.597(24)& 0.397(20)  \\
\hline
\end{tabular}
\caption{ Same as Table III for the form factor $ f_\perp$ in 
GeV$^{-1/2}$.
}
\end{center}
\end{table}

\section{ Chiral Extrapolations}
The form factors listed in Tables III and IV are for an unphysical 
world with $u$ and $d$ quark masses larger than in reality. 
They need to be extrapolated to the physical chiral limit. 
In ref.\cite{hpqcd} linear extrapolations were performed on the three 
points with $u_0 am_f = 0.01$, the only simulation data available 
at that time. 
Here we carry out this important step in several different ways : \\
1) linear extrapolation with only the full QCD ($m_f = m_q$) data, 
2) staggered chiral perturbation theory (SChPT) with full QCD data, 
3) continuum ChPT with full QCD data, and 
4) SChPT simultaneously to both full QCD and partially quenched data.
We use the last, most involved chiral extrapolation for our final 
answer, but make certain that the other methods give agreement within
quoted errors.

The formulas of chiral perturbation theory for heavy-light 
form factors have the general form,
\be
\label{schpt}
f_{\parallel/\perp} = c_0 [1 + \delta f_{\parallel/\perp} + c_1 m_q 
+ c_2 ( 2 m_f + m_s)  +  .... ] .
\ee
$\delta f_\parallel$ and $\delta f_\perp$ 
are the chiral log terms and they
have been determined at lowest order in $1/M$ for continuum, quenched and 
partially quenched QCD \cite{chpt1,chpt2} and also
specifically for staggered light quarks \cite{schpt1,schpt2,schpt3}. 
We use the formulas of 
 Aubin \& Bernard \cite{schpt3} that  include ``taste
 symmetry breaking'' lattice artifact terms 
that come in at ${\cal O}(a^2)$.  
$\delta f_\parallel$ and $\delta f_\perp$ 
 are functions of $E_\pi$ 
(more generally of $v \cdot p_\pi$, with $v^\mu$ equal to
 the $B$ four velocity and $p_\pi^\mu$ the pion momentum) and also 
depend on the $B^*B\pi$ coupling $g_{B\pi}$. 
 A possible 
source of concern in using ChPT may be the presence of pions with 
$E_\pi > 2 m_\pi$ in our system. Higher order terms in continuum heavy-light 
ChPT are discussed, for instance, in ref.\cite{stewart2}. 
 There it is argued that 
for some processes, effects of higher order terms can be taken into 
account by allowing for a correction to $g_{B\pi}$ linear in $E_\pi$. 
In our fits to SChPT and to continuum ChPT
 formulas we will let $g_{B\pi}$ vary as one
goes from one  $E_\pi$ value to another, together with $c_0$, $c_1$ 
and $c_2$.

The discussion in the previous paragraph shows that 
 chiral extrapolations are most conveniently carried out at fixed values 
of $E_\pi$. In order to do so, one needs to interpolate the data 
of Tables III \& IV to fixed common values of $E_\pi$ for each light 
quark mass.  We have explored  several different ways to perform the 
interpolations. We take advantage of various ansaetze that have been 
developed in the literature to model the $q^2$ dependence of form factors 
$f_+(q^2)$ and $f_0(q^2)$.  Appendix A summarizes commonly used ansaetze 
\cite{bk,bz,se1,se2,agrs,rh,bh}. 
The formulas given there refer to form factors in the real 
world, i.e. when using physical values for pion, $B$ and $B^*$ masses. 
We use them here also as guides for sensible interpolations 
even at unphysical meson masses.  We caution, however, that quantities 
such as $q^2_{max} = (M_B - m_\pi)^2$, $M_{B^*}^2 - q^2_{max}$ or the 
$B \, \pi$ threshold $(M_B + m_\pi)^2$ are sensitive to the pion mass. 
A particular ansatz may not work as well with unphysical masses 
compared to in the chiral limit, or vice versa.  For instance, 
we find that away from the chiral limit, i.e. when we are interpolating 
the data of Tables III and IV,  the most popular three parameter
Becirevic-Kaidalov (BK) ansatz 
\cite{bk} often fails to accommodate our data.  After the 
chiral extrapolation the BK ansatz works well, however. On the other hand we 
were often (though not always) able to fit data away from the chiral limit
 using just a simple single pole ansatz for both $f_0$ and $f_+$, 
but as expected,
a single pole ansatz does not work for $f_+(q^2)$ in the physical limit. 
The ansatz that works best both away and at the chiral limit is
 one due to Ball \& Zwicky (BZ)\cite{bz}, which actually is the same as the 
four parameter BK parametrization (see Appendix). Examples of fits 
used in the interpolations are given in Figs. 8 \& 9.  
  So, for each $(m_f,m_q)$ we first  convert the 
results for $f_\parallel$ and $f_\perp$ of Tables III \& IV to $f_0$ and $f_+$,
 interpolate using the BZ ansatz and finally convert back to 
$f_\parallel$ and $f_\perp$ at fixed values of $E_\pi$.  These are 
then extrapolated to the chiral limit using the SChPT formulas described 
above. One advantage of carrying out interpolations for $f_0$ and $f_+$ 
rather than directly in $f_\parallel$ and $f_\perp$, is that the 
kinematic constraint $f_0(0) = f_+(0)$ is easily incorporated into 
the ansaetze and one can use the $f_0$ data to constrain the 
normalization of $f_+$ as well.

\begin{figure}
\includegraphics[width=9.0cm,height=9.0cm,angle=-90]{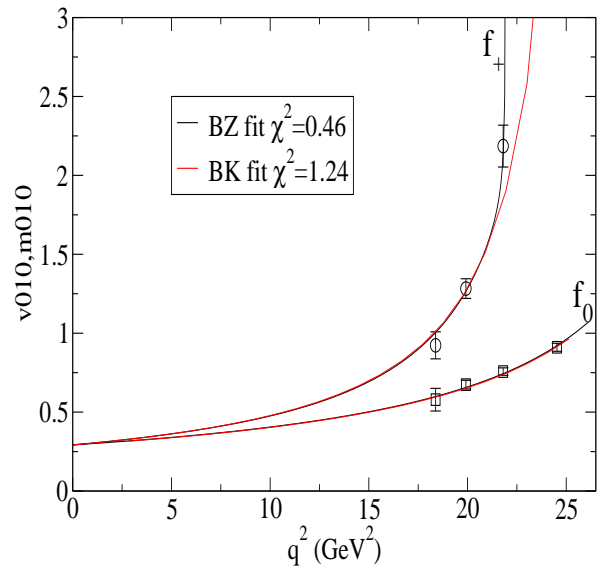}
\caption{ Comparison between a 3 parameter BK fit and a 4 parameter 
BZ fit to  $u_0am_q = u_0am_f = 0.01$ data.
 }
\end{figure}

\begin{figure}
\includegraphics[width=9.0cm,height=9.0cm,angle=-90]{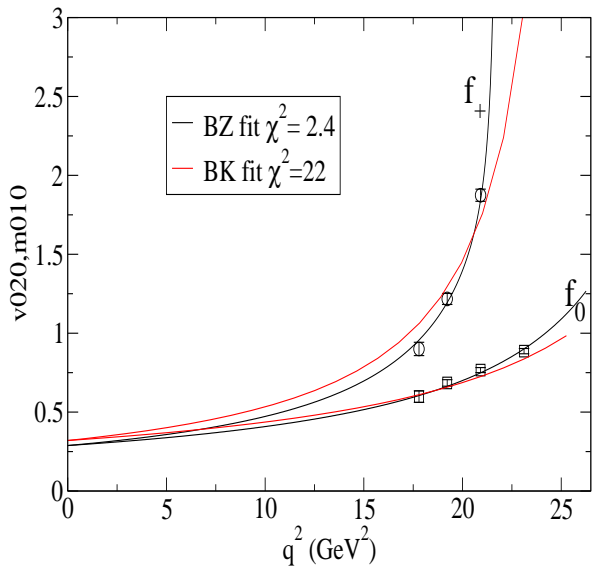}
\caption{ Same as Fig.8  for  $u_0am_q = 0.02,  \; u_0am_f = 0.01$.
 }
\end{figure}

In Figs.10 \& 11 we show chiral extrapolations for $f_\parallel$ and 
$f_\perp$ respectively for several values of $E_\pi$.  We compare 
results from simple linear extrapolations of full QCD data and from 
using the SChPT formulas simultaneously for both full QCD and partially 
quenched QCD data.  One sees that the difference between the two ways 
of performing the chiral extrapolations is small.  For $f_\parallel$ 
the central 
values at the chiral limit differ only by $ 0.1 \sim 2.4$\%.  For $f_\perp$ 
 the differences are again $< 2.4$\% for $E_\pi > 0.7$GeV.  For $E_\pi 
\leq 0.7$GeV (not shown in Fig.9) linear and SChPT extrapolations of $f_\perp$ 
 can differ by $4 \sim 6$\%.   On the other hand this is the
 kinematic region where 
statistical (and interpolation) errors are large for some 
combinations of $m_f$ and $m_q$ 
(see third column in Table IV) and it is not possible to 
disentangle statistical and chiral extrapolation errors. 
$f_\perp$ at small $E_\pi$ 
 translates into $f_+$ at large $q^2$.  One can speculate that the large 
statistical errors we are finding in this kinematic region may be due 
to the proximity of the $M_{B^*}$ pole.  This, however, for reasons 
we do not understand, was not evident in the partially quenched data 
 of reference \cite{hpqcd}.

 For the SChPT 
extrapolations, as mentioned above, we have let $g_{B \pi}$ float 
and be one of the fit parameters.  We find that the ``effective'' 
$g^2_{B\pi}$  ranges between 0.0 and 0.2 and 
decreases with $E_\pi$, although within large 
errors.  In Fig.12 we show a comparison of fits with and without 
the ${\cal O}(a^2)$ corrections in the ChPT theory formulas, i.e. 
a comparison between continuum and staggered full QCD ChPT. 
Again the differences are very small.  In summary, for most of our data points,
 different ways of carrying out the chiral extrapolation, 
including using no input from ChPT at all,  lead to a spread in 
extrapolated results of only $2.5$\% or less. 
 This indicates that contributions 
we have neglected in the ChPT formulas, such as $1/M$ corrections, 
higher order (in pion momentum) terms, finite volume effects etc. 
are  not important.  We take as central values for the form factors 
$f_\parallel$ \& $f_\perp$
in the chiral limit, the results coming from the SChPT extrapolation 
using both full QCD and partially quenched data. 
The combined statistical and chiral extrapolation errors are discussed 
in the next section.

\begin{figure}
\includegraphics[width=8.0cm,height=8.0cm]{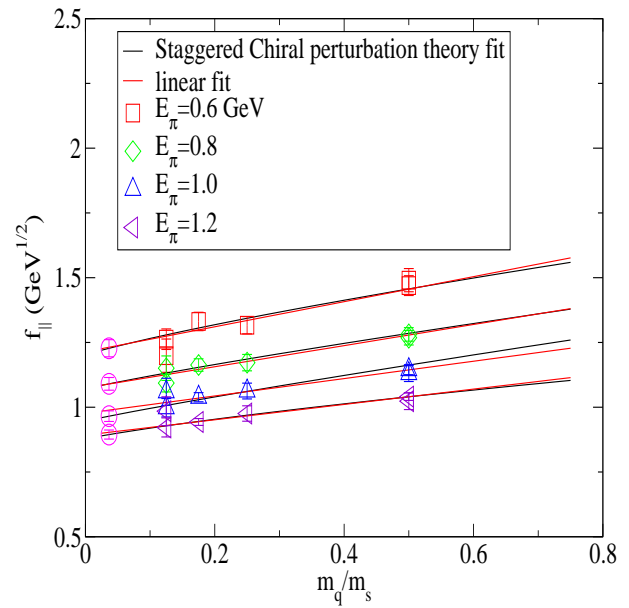}
\caption{ Chiral extrapolation of $f_\parallel$. 
The circles to the left show the extrapolated values at the 
physical limit.
 }
\end{figure}

\begin{figure}
\includegraphics[width=8.0cm,height=8.0cm]{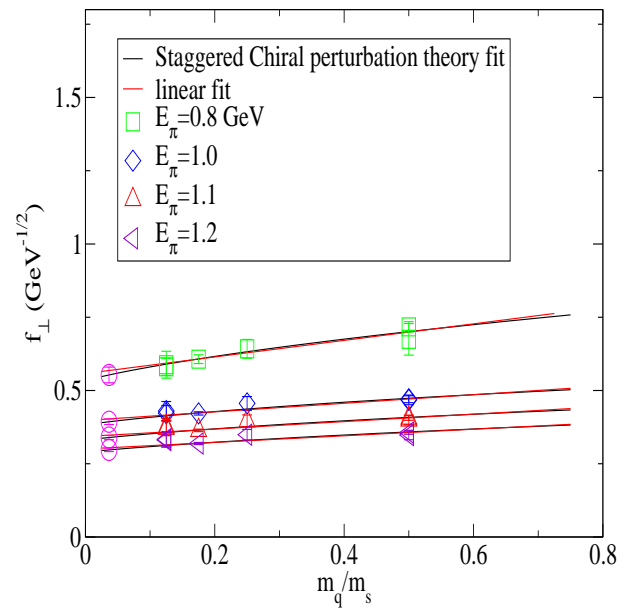}
\caption{ Same as Fig.8  for $f_\perp$. 
 }
\end{figure}

\begin{figure}
\includegraphics[width=8.0cm,height=8.0cm]{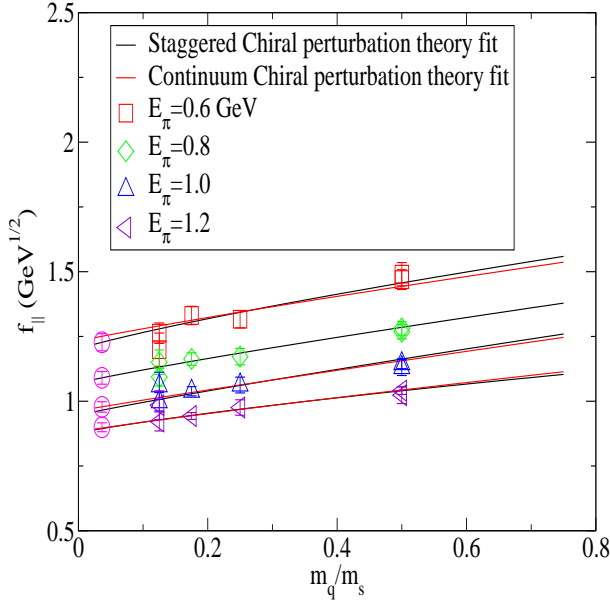}
\caption{ Chiral extrapolations with (staggered ChPT) and without
(continuum ChPT) ${\cal O}(a^2)$ corrections.  
 }
\end{figure}

\begin{figure}
\includegraphics[width=8.0cm,height=8.0cm]{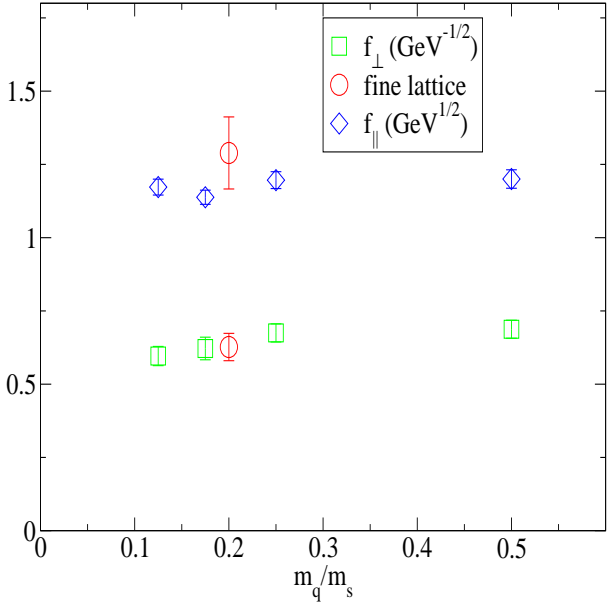}
\caption{ Comparison of coarse lattice data with  
some results from one of the fine MILC 
ensembles.  Shown are results for $f_\perp$ and $f_\parallel$
 at $E_\pi = 0.79$GeV.
 }
\end{figure}

In Fig.13 we compare results from one of the MILC fine lattice ensembles 
with the coarse lattice data discussed so far. For this comparison both 
the coarse and fine 
data points include only the $\langle J_\mu^{(0)} \rangle$ 
contributions through ${\cal O}(\alpha_s)$, i.e. the first lines in 
eqns.(\ref{v0}) or (\ref{vk}) respectively. Since we have shown that 
the higher order currents have minimal effect, we believe  
 meaningful scaling tests can be carried out
 with just $\langle J_\mu^{(0)} \rangle$.  For $f_\perp$, which is the main 
contributor to the phenomenologically relevant form factor $f_+(q^2)$, 
one sees that the fine lattice point falls nicely on the fixed $E_\pi$ 
curve determined by coarse lattice data. Our statistical error  
is large currently for the $f_\parallel$ point from the fine lattice.  
Within these large errors there is consistency between the coarse and 
fine lattices for $f_\parallel$ as well. We conclude from this exercise 
that there are no indications of large discretization effects in the 
form factor calculations on MILC coarse lattices. Such errors 
are smaller than current statistical errors.  Eventually it would be  
desirable to carry out a more thorough scaling test, once more data 
on fine lattices at several light sea quark masses become available.

\section{Results for Form Factors $f_+(q^2)$ and $f_0(q^2)$ in the 
Chiral Limit}
We convert the chirally extrapolated $f_\parallel(E_\pi)$ and 
$f_\perp(E_\pi)$ to the form factors $f_+(q^2)$ and 
$f_0(q^2)$ in the physical limit.  These are shown in Fig.14
and tabulated in Table V.   For comparison 
we also plot in Fig.14 the data presented  in ref.\cite{hpqcd}. One sees that 
changes are minimal in spite of all the improvements included in our new 
results.  This indicates that the approximations that were made previously and 
that we are systematically improving upon, such as partial quenching, 
linear chiral extrapolations, working 
with currents at lowest order in $1/M$,  did 
not drastically affect the theory.
The solid curves in Fig.14 are fits to our new results using
 the Ball-Zwicky (BZ) \cite{bz} parametrization of $f_+$ and $f_0$. 
We have also tried fits to other parametrizations, described in the Appendix, 
including the Becirevic-Kaidalov (BK) \cite{bk}, Richard Hill (RH) 
\cite{rh} and 
a series expansion (SE) \cite{se1,se2,bh,agrs} 
 parametrization. The RH parametrization fit 
is essentially indistinguishable from the BZ fit. The BK fit is also a 
good fit to our data although not quite as good as the first two. 
This should not be surprising, since the BK fit has only three parameters 
to tune whereas the BZ and RH fits are both
 four parameter fits.  Any further 
parameters, however, are very poorly determined and do not help in the fit. 
Another class of fit ansaetze, the series expansion (SE) fits, are discussed 
in the Appendix. 
 The main reason we are interested 
in obtaining a good analytic parametrization of 
the form factors, is to facilitate partial integration of 
differential decay rates, as discussed below.   These parametrizations 
can also be used to try and extrapolate to lower $q^2$  where 
lattice data are currently not available.

The statistical plus chiral extrapolation errors for $f_+(q^2)$ lie
between $7 \sim 10$\% depending on $q^2$.  They are smaller for 
the form factor 
$f_0(q^2)$.  For $q^2 \geq 16 GeV^2  $, the range we will be 
focusing on, the average error for $f_+(q^2)$ comes out to be $\sim8$\%.  
In Table VI we list this average statistical plus chiral extrapolation error 
together with estimates of systematic errors from other 
sources. These other systematic errors are dominated 
by the $\sim 9$\% uncertainty in higher order matching of the heavy-light 
currents.

\begin{figure}
\includegraphics[width=7.5cm,height=7.5cm]{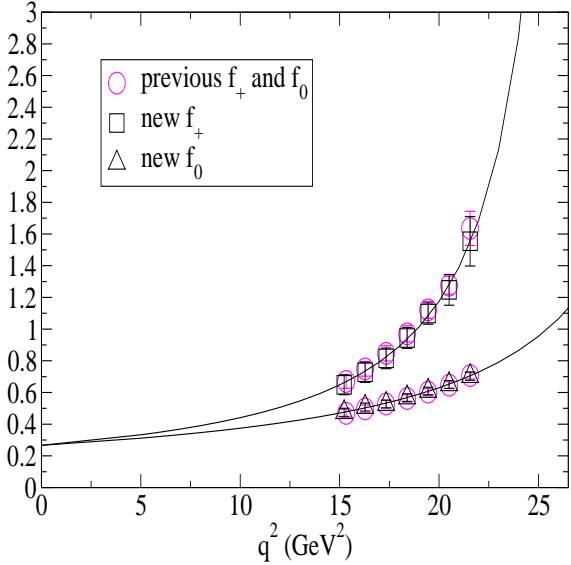}
\caption{ Form factors $f_+(q^2)$ and $f_0(q^2)$ in the chiral limit. 
The black squares and triangles are the new and final results 
for $f_+$ and $ f_0$ respectively. 
For comparison, the data from ref.\cite{hpqcd} are also shown as circles.
The full black curves follow
  a BZ parametrization fit (see text) to the new data.
Errors are combined statistical and chiral extrapolation errors. 
 }
\end{figure}

The differential partial
 decay rate for $B \rightarrow \pi\, l \nu$, ignoring the
 charged lepton mass, is given by,
\be
 \frac{d\Gamma}{dq^2} = \frac{G_F^2 }
{24 \pi^3 } \, p_\pi^3 \,  |V_{ub}|^2 \;  |f_+(q^2)|^2
\ee
where $G_F$ is the Fermi constant and $p_\pi$ the magnitude of the 
pion three momentum in the $B$ rest frame. 
Knowing $f_+(q^2)$ then allows us to evaluate $\frac{1}{|V_{ub}|^2} 
\, \frac{d \Gamma}{d q^2}$ and also integrate this 
quantity over different $q^2$ bins. We take our best fit, the BZ fit shown in 
Fig.14, and integrate to obtain,
\be
\label{diffg}
\frac{1}{|V_{ub}|^2} \, \int_{16 GeV^2}^{q^2_{max}} \frac{d\Gamma}{d q^2} \,
dq^2 = 1.46(23)(27) \,  ps^{-1}
\ee
  The first error is 
the combined statistical plus chiral extrapolation error and the second 
the sum of all other systematic errors added in quadrature.  
Eq.(\ref{diffg}) is the main result of this article. It serves as basis for
determinations of the CKM matrix element $|V_{ub}|$.  
Similar integrals using other parametrizations and over other $q^2$ ranges 
are summarized in Table VII together with 
 $f_+(0)$ from the different fits.
The second error for $f_+(0)$ and for 
 the integrated rate over the entire $q^2$ range includes an 
additional $10$\% systematic uncertainty in $f_+(q^2)$ which is not 
part of Table VI and which 
comes from the extrapolation into the low $q^2$ region.

\begin{table}
\begin{center}
\begin{tabular}{|c|c|c|}
\hline
$ q^2 $ [GeV$^2$]  &  $\qquad f_+(q^2) \qquad$  &  $ \qquad f_0(q^2)
\qquad $  \\
\hline
\hline
15.23 & 0.649(63) & 0.475(26) \\
16.28 & 0.727(64) & 0.508(25) \\
17.34 & 0.815(65) & 0.527(25) \\
18.39 & 0.944(66) & 0.568(24) \\
19.45 & 1.098(67) & 0.610(24) \\
20.51 & 1.248(97) & 0.651(25) \\
21.56 & 1.554(156) & 0.703(26) \\
\hline
\end{tabular}
\caption{ Form factors $f_+(q^2)$ and $f_0(q^2)$ in the chiral limit.
Errors shown are combined statistical and chiral extrapolation errors.
}
\end{center}
\end{table}

\begin{table}
\begin{center}
\begin{tabular}{|c|c|}
\hline
source of error & size of error (\%) \\
\hline
\hline
statistics + chiral extrapolations  &  8  \\
two-loop matching   &                  9  \\
discretization      &                  3  \\
relativistic        &                  1   \\
\hline
  Total             &                 12  \\
\hline
\end{tabular}
\caption{ Estimate of percentage errors in $f_+(q^2)$ 
for $q^2 > 16$GeV$^2$. 
}
\end{center}
\end{table}

\begin{table}
\begin{center}
\begin{tabular}{|c|c|c|c|}
\hline
Fit  &  $f_+(q^2=0)$  &
\multicolumn{2}{c|} {$ \int \frac{d \Gamma}{d q^2} / |V_{ub}|^2$ [$ 
\, ps^{-1}$]} \\
\hline 
 & & $\; 0 \leq q^2 \leq q^2_{max} \;$ & 
 $ 16 GeV^2 \leq q^2 \leq q^2_{max}$ \\
\hline
\hline
BZ  & 0.27(2)(4) & 6.00(96)(1.68) &  1.46(23)(27)  \\
BK  & 0.26(2)(4) & 6.03(96)(1.69) &  1.31(21)(25)  \\
RH  & 0.27(2)(4) & 5.99(96)(1.68) &  1.45(23)(27) \\
\hline
\end{tabular}
\caption{ Partially integrated differential decay rates and $f_+(0)$
using several parametrizations.  The first error reflects statistical 
and chiral extrapolation uncertainties.  The second error is due to 
remaining systematic errors.
}
\end{center}
\end{table}

\section{ Estimating $|V_{ub}|$}
In this section we combine the lattice results of the previous section with 
experimental input for $B \rightarrow \pi \, l \nu$ branching fractions and 
extract an estimate for $|V_{ub}|$.  In order to do so, we use 
 results from the Heavy Flavor Averaging Group (HFAG) \cite{hfag} 
and rely on its analysis of the current experimental uncertainties.
The  HFAG gives  preliminary averages of BaBar, Belle and CLEO results 
(as of the conferences of Summer 2005) 
for the integrated branching fraction 
${\cal B} \{B^0 \rightarrow \pi^- \, l^+ \nu \}$.  They quote 
 $[1.35 \pm 0.08 \pm 0.08] \times 10^{-4}$ for
$0 \leq q^2 \leq q^2_{max}$ and $[0.40 \pm 0.04 \pm 0.04] \times 10^{-4}$ for 
$q^2 \geq 16$GeV$^2$.  Combining this with eq.(\ref{diffg}) and a 
$B^0$ lifetime of $1.536 \, ps$ \cite{pdg} leads to,
\be
\label{vub1}
|V_{ub}| = 4.22(30)(51) \times 10^{-3},  \qquad q^2 \geq 16 GeV^2
\ee
where the first error is experimental (7\%) and the second is the 
total lattice error (12\%). The result (\ref{vub1}) is consistent, 
at the one $\sigma$ level, 
with the preliminary value $3.78(25)(52) \times 10^{-3}$ 
obtained by the Fermilab/MILC collaboration using 
the same HFAG branching fraction averages \cite{okamoto}.

\section{Summary}
We have completed a determination of the $B$ meson semileptonic 
form factors $f_+(q^2)$ and $f_0(q^2)$ using state-of-the-art 
Lattice QCD methods. Our calculations employ unquenched gauge configurations, 
created by the MILC collaboration, that incorporate vacuum polarization 
effects from two very light flavors plus strange sea quarks.
 Both the sea and 
the valence light quarks are simulated using a highly improved staggered quark 
action.  This action allows us to work close enough to the chiral limit, 
so that chiral extrapolations to physical pions are mild and do not 
introduce large uncertainties. Our results for $f_+(q^2)$ can be 
combined with experimental branching fraction data 
to extract the CKM matrix element $|V_{ub}|$.  This quantity is a 
crucial ingredient in tests of the unitarity triangle and in 
solidifying our understanding of CP violation in the Standard Model.
 
The total lattice error in the $f_+(q^2)$ form factor
 presented here, and hence also in 
$|V_{ub}|$, is at the $\sim 12$\% level. This error is dominated 
by uncertainty in higher order perturbative matching of heavy-light 
currents and by statistical errors.  One of the 
 goals of the HPQCD collaboration is to carry out higher order matching 
for heavy-light, light-light and heavy-heavy currents and four-fermion 
operators \cite{howard}. 
It has become increasingly obvious that such calculations 
are necessary for accurate lattice determinations of form factors, decay 
constants and mixing parameters at the $\sim 5$\% level. 
For form factor calculations, more work is also required to reduce 
statistical errors.  One approach that helped significantly in the HPQCD 
collaboration $f_B$ determination \cite{fb} is to explore different 
smearings of sources and sinks in correlators. We plan to investigate 
this in the future.

Lattice results for $f_+(q^2)$ exist at the moment only for $q^2 \geq 
16GeV^2$ and this forces us to use only part of the available experimental 
branching fraction data.  In order to take advantage of all the 
experimental data and thereby reduce experimental errors in 
$|V_{ub}|$,  other methods such as sum rules approaches 
are currently employed \cite{bz1,bz}
 to cover the low $q^2$ region.  A lattice approach 
to $f_+(q^2)$ determinations at low $q^2$, ``Moving NRQCD'', 
 was introduced many years ago
and work is in progress to implement this method using our 
highly improved quark and gauge actions \cite{mnrqcd1,mnrqcd2}.
We can look forward to 
the next major improvement in lattice determinations of $|V_{ub}|$ 
once two-loop matching, reduction in statistical errors 
and control over the entire $q^2$ range has been achieved.

\vspace{.3in}
\noindent
\underline{Acknowledgements}

\vspace{.1in}
\noindent
This work was supported by the DOE and NSF (USA) and by PPARC (UK).
 A.G., J.S. and M.W. thank the KITP 
U.C. Santa Barbara for support during the workshop,
 ``Modern Challenges in Lattice Field Theory'' when part of the present 
research was carried out.
Simulations were done at NERSC and on the 
Fermilab LQCD cluster.  We thank the MILC collaboration 
for making their unquenched gauge configurations available 
 and the Fermilab collaboration for use of 
their light propagators on the fine lattices.
We are grateful to Claude Bernard 
for sending us his notes on SChPT for heavy-light form factors.
We would also like to acknowledge 
useful conversations with Richard Hill, 
Masataka Okamoto and Iain Stewart.

\appendix

\section{Parametrization of Form Factors}
Most parametrizations start from a dispersive representation of
the form factors.
\begin{eqnarray}
f_+(q^2) & = & \frac{r_1}{(1 -\tilde{q}^2)} + \frac{1}{\pi}
\int_{t_+}^\infty dt \frac{Im[f_+(t)]}{t-q^2-i \epsilon} \nonumber \\
 & &  \nonumber \\
&\longrightarrow &
 \frac{r_1}{(1 -\tilde{q}^2)} +
\frac{r_2}{(1 - \alpha \,\tilde{q}^2)}   \\
 & &  \nonumber \\
f_0(q^2) & \longrightarrow & \frac{r_1 + r_2}{(1 -
\tilde{q}^2/\beta)} 
\end{eqnarray}
\noindent
where  $\tilde{q}^2 \equiv q^2/M^2_{B*}$
 and  $t_+ \equiv (M_B + m_\pi)^2$.
The kinematic constraint $f_+(0) = f_0(0)$ is automatically satisfied.
This is a 4 parameter parametrization of $f_+$ and $f_0$, sometimes
called the ``4 Parameter BK Parametrization'' (Becirevic \& Kaidalov)
\cite{bk}. 
Examples of generalizations and special cases developed in the literature 
are given below. They differ only in the parametrization of $f_+$.

\vspace{.1in}
\noindent
\underline{3 Parameter Becirevic-Kaidalov (BK) \cite{bk} }
\begin{eqnarray}
f_+(q^2) &=& \frac{f_+(0)}{(1-\tilde{q}^2) \, (1 - \alpha
\tilde{q}^2)},  \\
f_0(q^2) &=& \frac{f_+(0)}{(1 - \tilde{q}^2/\beta)}. \\
&&  \nonumber   \\
&&  \nonumber 
\end{eqnarray}

\noindent
\underline{ 4 Parameter Ball-Zwicky (BZ) \cite{bz}}
\begin{eqnarray}
f_+(q^2) &=& \frac{f_+(0)}{(1-\tilde{q}^2)} + \frac{r \tilde{q}^2}
{(1-\tilde{q}^2)(1 - \alpha \tilde{q}^2)}. \\
&&  \nonumber   \\
&&  \nonumber 
\end{eqnarray}

\noindent
\underline{ 4 Parameter R.Hill (RH) \cite{rh}}
\be
f_+(q^2) = \frac{f_+(0) (1 - \delta \cdot \tilde{q}^2)}
{(1- \tilde{q}^2)(1 -  \tilde{q}^2/\gamma)} .
\ee

\vspace{.1in}
\noindent
Another class of parametrizations, advocated in \cite{se1,se2,agrs} and also 
discussed in \cite{bh}, is based on a series expansion of $f_+(q^2)$ 
around some $q^2 = t_0$.  For better convergence of the series it is customary 
to convert to a new variable $z(q^2,t_0)$, where following \cite{agrs,bh}
 we take,
\be
z(q^2,t_0) = \frac{\sqrt{t_+ - q^2} - \sqrt{t_+ - t_0}}
                 {\sqrt{t_+ - q^2} + \sqrt{t_+ - t_0}}.
\ee
We refer to the literature for further discussion of the merits of 
this transformation. The form factor $f_+(q^2)$ can then be expanded 
as a power series in $z(q^2,t_0)$,
\be
f_+(q^2) = \frac{1}{P(q^2) \, \Phi(q^2,t_0)} \; \sum_{k=0}^{k_{max}} 
a_k(t_0) \, [z(q^2,t_0)]^k.
\ee
The ``Blaschke'' factor $P(q^2)$ must take into account any isolated poles
 below the $B \, \pi$ threshold at $q^2 = t_+$. We set $P(q^2) =
z(q^2,M_{B^*}^2)$ to take care of the $B^*$ pole. For $\Phi(q^2,t_0)$ 
we take the expression given in ref.\cite{agrs}
 (with simplified $\chi^{(0)}$). We 
combine the ansatz (A8) for $f_+(q^2)$ with $k_{max} = 2$ together with
 (A4) for 
$f_0(q^2)$ to get another (and our last) 4 parameter ansatz, \\

\vspace{.1in}

\noindent
\underline{ Series Expansion (SE)}
\be
f_+(q^2) = \frac{ 
 a_0 + a_1 z(q^2,t_0) + a_2 z^2 }
{z(q^2,M_{B^*}^2) \, \Phi(q^2,t_0)} .
\ee

\noindent
We have explored the SE parametrization but not because we needed a
better analytic expression to cover the range $q^2 \geq 16GeV^2$ of our
lattice results.  Rather, we did so to assess how reliable is
the information on 
 the shape of the form factors at lower $q^2$ that we are getting 
from the BK/BZ/RH type parametrizations. 
We find that good fits to the simulation data 
can be obtained with the SE parametrizations as well 
and that they are insensitive to the value of $t_0$. 
Table VIII gives results for integrated differential 
decay rates.  Not surprisingly, agreement is found with Table VII for 
$q^2 \geq 16GeV^2$. Results are systematically slightly lower than in 
Table VII if one includes the entire $q^2$ range. One can summarize by saying 
that we do not have evidence for any strong dependence on the choice of 
ansatz employed to extrapolate lattice data to lower 
$q^2$ values.  Nevertheless, 
we believe it is important  to look for
additional information on the form factors at $q^2 < 16GeV^2$, either from 
experiment or from theoretical models, if one wants to discuss the entire
$q^2$ range at the present time.
This  has already been done in the 
recent literature \cite{bh,agrs,bz}. 
 Alternatively, getting lattice results directly in the 
low $q^2$ regime using for instance Moving NRQCD would also solve this 
problem.

\begin{table}
\begin{center}
\begin{tabular}{|c|c|c|}
\hline
 $t_0/q^2_{max}$  &
\multicolumn{2}{c|} {$ \int \frac{d \Gamma}{d q^2} / |V_{ub}|^2$ [$ 
\, ps^{-1}$]} \\
\hline 
 & $\; 0 \leq q^2 \leq q^2_{max} \;$ & 
 $ 16 GeV^2 \leq q^2 \leq q^2_{max}$ \\
\hline
\hline
0.0     &  5.78    & 1.32  \\
0.2     &  5.78    & 1.33  \\
0.4     &  5.79    & 1.33  \\
0.6     &  5.80    & 1.33  \\
0.8     &  5.79    & 1.33  \\
1.0     &  5.79    & 1.33  \\
\hline
\end{tabular}
\caption{ Partially integrated differential decay rates using 
the SE parametrization for different choices of $t_0$.  
No error estimates have been made.
}
\end{center}
\end{table}

\section{Erratum:  B Meson Semileptonic Form Factors 
from Unquenched  Lattice QCD [Phys.Rev.D {\bf 73}, 074502 (2006)]}

\noindent
Due to a normalization error in one of our analysis codes 
many  results for the form factor $f_\perp$ were evaluated 
incorrectly. In particular, Table IV should be replaced by,
\begin{center}
\begin{tabular}{|c|c|c|c|c|}
\hline
\multicolumn{5}{|c|}{ $f_\perp$ [Gev$^{-1/2}$]} \\
\hline
$u_0 a m_f$ & $u_0 am_q$  & $p_\pi = (001)$  & 
$ (011)$ & $ (111)$\\
\hline
\hline
0.005  & 0.005  &1.543(205) &0.994(62) & 0.892(68)  \\
0.007  & 0.007  &1.082(31) &0.857(44) & 0.750(83)  \\
0.010  & 0.005  &1.128(37) &0.936(51) & 0.715(66)  \\
0.010  & 0.010  &1.235(90) &0.929(52) & 0.778(74)  \\
0.010  & 0.020  &1.029(21) &0.864(27) & 0.733(40)  \\
0.020  & 0.020  &1.097(140) &0.844(34)& 0.688(35)  \\
\hline
\end{tabular}
\end{center}

\vspace{.2in}
\noindent
The corrected form factors $f_+(q^2)$ and $f_0(q^2)$ in the chiral 
limit become (previous Table V):
\begin{center}
\begin{tabular}{|c|c|c|}
\hline
$ q^2 $ [GeV$^2$]  &  $\qquad f_+(q^2) \qquad$  &  $ \qquad f_0(q^2)
\qquad $  \\
\hline
\hline
17.34 & 1.101(53) & 0.561(26) \\
18.39 & 1.273(99) & 0.600(21) \\
19.45 & 1.458(142) & 0.639(23) \\
20.51 & 1.627(185) & 0.676(41) \\
21.56 & 1.816(126) & 0.714(56) \\
\hline
\end{tabular}
\end{center}
We restrict the range in $q^2$ to values where simulation data before 
chiral extrapolation are available for all light quark masses 
 and omit two lower $q^2$ points 
that were obtained previously through small extrapolations in $E_\pi$.

\vspace{.2in}
\noindent
The new error budget is given by (previous Table VI):
\begin{center}
\begin{tabular}{|c|c|}
\hline
source of error & size of error (\%) \\
\hline
\hline
statistics + chiral extrapolations  &  10  \\
two-loop matching   &                  9  \\
discretization      &                  3  \\
relativistic        &                  1   \\
\hline
  Total             &                 14  \\
\hline
\end{tabular}
\end{center}

\vspace{.2in}
\noindent
and the partially integrated differential decay rates become 
(previous Table VII):
\begin{center}
\begin{tabular}{|c|c|c|c|}
\hline
Fit  &  $f_+(q^2=0)$  &
\multicolumn{2}{c|} {$ \int \frac{d \Gamma}{d q^2} / |V_{ub}|^2$ [$ 
\, ps^{-1}$]} \\
\hline 
 & & $\; 0 \leq q^2 \leq q^2_{max} \;$ & 
 $ 16 GeV^2 \leq q^2 \leq q^2_{max}$ \\
\hline
\hline
BZ  & 0.31(5)(4) & 9.10(1.82)(2.55) &  2.07(41)(39)  \\
BK  & 0.31(5)(4) & 9.30(1.86)(2.60) &  2.13(43)(40)  \\
SE  & 0.30(5)(4) & 9.35(1.87)(2.62) &  2.02(40)(38)  \\
\hline
\end{tabular}
\end{center}

\vspace{.2in}
\noindent
Equation (22) should be replaced by,

$$ 
\qquad
\frac{1}{|V_{ub}|^2} \, \int_{16 GeV^2}^{q^2_{max}} \frac{d\Gamma}{d q^2} \,
dq^2 = 2.07(41)(39) \,  ps^{-1},
$$
 
\noindent
which signifies an increase on our previous
 incorrect value ( 1.46(23)(27) $ps^{-1}$ ) by 1.7 times the previously 
quoted total error.

\vspace{.2in}
\noindent
The new eq.(22) leads to a new 
 eq.(23), 

$$
|V_{ub}| = 3.55(25)(50) \times 10^{-3},  \qquad q^2 \geq 16 GeV^2.
$$

\vspace{.2in}
\noindent
Accordingly, the last two sentences in the Abstract should be changed to,\\
``.....\\
We calculate the form factors
$f_+(q^2)$  and $f_0(q^2)$ in the chiral limit for the range $16 \,
 {\rm GeV}^2 \leq q^2 < q^2_{max}$ and obtain  $
 \int_{16 GeV^2}^{q^2_{max}} [\,d\Gamma/d q^2\,] \,
dq^2 \; / \;|V_{ub}|^2 = 2.07(57) \, ps^{-1}$.
Combining this with a preliminary average by the Heavy Flavor Averaging Group
(HFAG'05) of recent branching fraction data 
for exclusive B semileptonic decays from the BaBar, Belle and CLEO
collaborations, leads to $|V_{ub}| = 3.55(25)(50) \times 10^{-3}$.''

\newpage


\begin{thebibliography}{99}
\bibitem{ups2}
The NRQCD action used in this article is indeed the same as in \cite{upsilon}, 
which, however, included some typos in the $c_2$ and $c_3$ terms of $\delta H$.
Those typos have been corrected here in eq.(3).
\end{thebibliography}

\end{document}